\documentclass[preprint,12pt]{elsarticle}

\usepackage{graphicx}
\usepackage{amssymb}
\usepackage{amsmath}
\usepackage{rotating}
\usepackage{url}

\newcommand{\bra}[1]{\left<#1\right|}
\newcommand{\ket}[1]{\left|#1\right>}
\newcommand{\bket}[2]{\left<#1|#2\right>}
\newcommand{\braket}[3]{\bra{#1}#2\ket{#3}}

\journal{Journal of Computational Physics}

\begin{document}

\begin{frontmatter}


\title{\texttt{mudirac}: a Dirac equation solver for elemental analysis with muonic X-rays}


\author{Simone Sturniolo}
\ead{simone.sturniolo@stfc.ac.uk}
\address{Scientific Computing Department, UK Research and Innovation, Harwell Campus, Didcot OX11 0GD (UK)}

\begin{abstract}
We present a new open source software for the integration of the radial Dirac equation developed specifically with muonic atoms in mind. The software, called \texttt{mudirac}, is written in C++ and can be used to predict frequencies and probabilities of the transitions between levels of the muonic atom. In this way, it provides an invaluable tool in helping with the interpretation of muonic X-ray spectra for elemental analysis. We introduce the details of the algorithms used by the software, show the interpretation of a few example spectra, and discuss the more complex issues involved with predicting the intensities of the spectral lines. The software is available publicly at \url{https://github.com/muon-spectroscopy-computational-project/mudirac}.
\end{abstract}


\end{frontmatter}


\section{Introduction}\label{S:intro}

Muonic atoms, in which one electron is replaced by the much heavier muon, have long been known and studied as an interesting laboratory for many kinds of exotic effects and insights in physics \cite{borie1982}. Thanks to its higher mass, a muon surrounding an atomic nucleus has orbitals much smaller than an electron, and thus interacts far more strongly with the nucleus, probing effects such as vacuum polarisation \cite{dixit1971} and the internal structure of the proton \cite{pachucki1999}. More recently, interest in using muonic atoms for elemental analysis has arisen \cite{hillier2016,ninomiya2010}, with experiments carried out at international muon facilities such as J-PARC and ISIS. Muons are implanted in a sample with a beam, and then the X-rays emitted during their cascade are monitored and used to determine its composition. In this way, muons provide a very powerful probe that's non-destructive and can be tuned in depth by varying the momentum of the incident beam. This technique is especially useful when analysing archaeological artifacts, which could be damaged by any approach that required extracting a sample from them.\newline
While there exist tabulated databases of muonic atom X-ray frequencies \cite{zinatulina2018} and dedicated codes have been developed in the past to compute them \cite{rinker1979}, we still regarded it as useful to develop a new, modern code that could be used to aid with muon elemental analysis at ISIS. With this in mind, we have developed \texttt{mudirac}, an open source C++ solver for the radial Dirac equation, designed specifically for computation of muonic X-ray spectra. \texttt{mudirac} can compute muonic atom eigenenergies up to precision of less than 1 keV, accounting for the effects of finite nuclear size, vacuum polarizability, and electronic shielding, predicting both energies and transition probabilities for X-ray lines. This paper will provide an overview of the algorithms employed, a short documentation on how to use the software, and finally some tests and comparisons with reference sample spectra.

\section{Theory}\label{S:theory}

\subsection{The radial Dirac equation}\label{S:theory:dirac}

The Dirac equation for a radial potential can be reduced to a pair of first order one dimensional differential equations \cite{gross1999,weinb2008,silbar2010}:

\begin{align}\label{s:theory:dirac:dirac}
	\frac{\partial Q}{\partial r} &= \frac{k}{r}Q + \left(mc-\frac{E-V(r)}{c}\right)P \\
	\frac{\partial P}{\partial r} &= -\frac{k}{r}P + \left(mc+\frac{E-V(r)}{c}\right)Q \nonumber
\end{align}

in which $P$ and $Q$ represent two components of the equation, with $P$ being the `large' one when close to the non-relativistic regime. Here $m$ is the mass of the orbiting particle, or can be replaced by the effective mass, $\mu = (1/m+1/M)^{-1}$, including the effect of recoil from the nuclear mass $M$ if one does not treat the nucleus as having infinite mass. In this equation and all the following atomic units are used, so that $e = \hbar = 1$ and $c = 1/\alpha \sim 137$. $k$ is a quantum number that can take on any positive or negative integer values (but not zero). In terms of the angular momentum and total angular momentum quantum numbers used in the Schr\"{o}dinger equation, $l$ and $j$, we have that

\begin{equation}\label{s:theory:dirac:kqnum}
k = 
    \begin{cases}
    -(l+1) \qquad &j = l+1/2 \\
    l \qquad &j = l-1/2 
    \end{cases}
\end{equation}
. The full three-dimensional wave function can then be reconstructed as:

\begin{equation}\label{s:theory:dirac:fullpsi}
    \bket{\mathbf{r}}{\psi_{k\mu}} = 
    \begin{pmatrix}
        \frac{P_k}{r}\bket{\mathbf{\hat{r}}}{k\mu} \\
        i\frac{Q_k}{r}\bket{\mathbf{\hat{r}}}{-k\mu}
    \end{pmatrix}
\end{equation}

where we make use of the spin spherical harmonics:

\begin{equation}\label{s:theory:dirac:spsphharm}
\ket{k\mu} = \sum_{s=\pm 1/2} c(l,\frac{1}{2};\mu-s, s)\left|l,\mu-s\right>\Phi(s)
\end{equation}

with 

\begin{align}\label{s:theory:dirac:sphharm}
    \left<\mathbf{\hat{r}}|l,\mu-s\right> = Y_{l,\mu-s}(\mathbf{\hat{r}}) \\
    \Phi\left(\frac{1}{2}\right) = \begin{pmatrix}
    1 \\
    0
    \end{pmatrix}
    \qquad
    \Phi\left(-\frac{1}{2}\right) = \begin{pmatrix}
    0 \\
    1
    \end{pmatrix} 
\end{align}

, $Y_{lm}$ being the usual spherical harmonics, and the Clebsch-Gordan coefficients

\begin{equation}\label{s:theory:dirac:clebschgordan}
c(l,\frac{1}{2};\mu-s, s) = \begin{cases}
-\mathrm{sgn}(k)\sqrt{\frac{k+\frac{1}{2}-\mu}{2k+1}} \qquad s = \frac{1}{2} \\
\sqrt{\frac{k+\frac{1}{2}+\mu}{2k+1}} \qquad s = -\frac{1}{2}
\end{cases}
\end{equation}

as seen in \cite{gross1999,biedernharn1981}, and remembering eq. \ref{s:theory:dirac:kqnum} for the relationship between $k$ and $l$.\newline
Particular attention should also be paid to nodal theorems for this equation, as they are a useful guide when searching for solutions. The rule is that the large component of the wavefunction, $P$, has always $n-l-1$ nodes, while the $Q$ component has $n-l-1$ nodes if $k < 0$, and $n-l$ nodes if $k > 0$ \cite{hall2014, johnson2007}.

\subsection{Potential}\label{S:theory:pot}

The base potential used in eq. \ref{s:theory:dirac:dirac} is the Coulomb potential, with an assumption of the nucleus being a uniformly charged sphere if a finite radius is used:

\begin{equation}\label{s:theory:pot:coulomb}
    V(r) =
 \begin{cases}
-\frac{Z}{R}\left(\frac{3}{2}-\frac{1}{2}\frac{r^2}{R^2}\right), & \text{$r\leq R$}.\\
-\frac{Z}{r}, & \text{$r > R$}.
\end{cases}
\end{equation}

. The radiuses of most isotopes of physical interest are extracted from the work by Angeli and Marinova \cite{angeli2013}.  It should be noted that the values tabulated in the paper are values for the root mean square radius, which means they need to be multiplied by $\sqrt{5/3}$ to get the equivalent radius of a uniformly charged sphere. For any isotope for which the radius is not known, we fall back on the empirical formula \cite{borie1982}:

\begin{equation}\label{nuc_radius}
R = 1.2 A^{1/3} \mathrm{fm} = 2.267\cdot10^{-5}A^3
\end{equation}

, where the second version is in atomic units.\newline
Two possible corrections to the Coulomb potential have been considered relevant for this software tool, as they've been shown to be largely the biggest contributions to transition energies, especially in heavy muonic atoms \cite{borie1982}. The first one is the so-called Uehling potential, representing the first order contribution of vacuum polarisability to the interaction between the nucleus and the muon - in other words, the contribution of a Feynman diagram in which a photon exchange between muon and nucleus is mediated through an electron-positron pair production loop. This potential has a well known form and can be written as

\begin{equation}\label{s:theory:pot:uehling-point}
V_U(r) = -\frac{2\alpha Z}{3\pi r}\int_0^1 du \sqrt{1-u^2}\left(\frac{1}{u}+\frac{1}{2}u\right)e^{-2cr/u}
\end{equation}

for a point charge \cite{uehling1935,wichmann1956}, and 

\begin{equation}\label{s:theory:pot:uehling-distr}
V_U^{(distr)}(r) = -\frac{2\alpha^2}{3 r}\int_0^\infty dx x \rho(x)\int_0^1 du \sqrt{1-u^2}\left(1+\frac{1}{2}u^2\right)\left(e^{-2c|r-x|/u}+e^{-2c(r+x)/u}\right)
\end{equation}

for a generic charge distribution $\rho(r)$ \cite{fullerton1976}. For the specific case of a uniform spherical charge density,

\begin{equation}\label{s:theory:pot:rho}
\rho(r) = \begin{cases}
\frac{3}{4\pi}\frac{Z}{R^3} \qquad &r \leq R \\
0 \qquad &r > R \\
\end{cases}
\end{equation}

the integral becomes

\begin{equation}\label{s:theory:pot:uehling-sphere}
V_U^{(sphere)}(r) = -\frac{\alpha^2Z}{2\pi rR^3}\int_0^R dx x \int_0^1 du \sqrt{1-u^2}\left(1+\frac{1}{2}u^2\right)\left(e^{-2c|r-x|/u}+e^{-2c(r+x)/u}\right)
\end{equation}

. The integral in $x$ can be carried out if we consider the two distinct cases of $x < r$ and $x \geq r$. For the first we have

\begin{multline}\label{s:theory:pot:intx_less}
X_{<}(r, u | R) = \int_0^R x \left(e^{-2c|r-x|/u}-e^{-2c(r+x)/u}\right)  dx = \\
e^{-2cr/u}\int_0^R x \left(e^{2cx/u}-e^{-2cx/u}\right)  dx = \\ e^{-2cr/u}\left[e^{2cR/u}\left(\frac{Ru}{2c}-\frac{u^2}{4c^2}\right)+e^{-2cR/u}\left(\frac{Ru}{2c}+\frac{u^2}{4c^2}\right)\right]
\end{multline}

while for the second

\begin{multline}\label{s:theory:pot:intx_more}
X_{>}(r, u | R) =\int_0^R x \left(e^{-2c|r-x|/u}-e^{-2c(r+x)/u}\right)  dx = \\
\int_0^R x \left(e^{-2c(x-r)/u}-e^{-2c(x+r)/u}\right)  dx = \\ \left(e^{2cr/u}-e^{-2cr/u}\right)\left[-\frac{Ru}{2c}e^{-2Rc/u}-\frac{u^2}{4c^2}(e^{-2Rc/u}-1)\right]
\end{multline}

. It then follows that the full term can be expressed as

\begin{multline}\label{s:theory:pot:uehling-sphere-int}
V_U^{(sphere)}(r) =  -\frac{\alpha^2Z}{2\pi rR^3}\int_0^1 du \sqrt{1-u^2}\left(1+\frac{1}{2}u^2\right) \times \\
\left[X_{<}(r, u | \min(r, R)) + X_{>}(r, u | R) -X_{>}(r, u | \min(r, R))\right]
\end{multline}

which is the expression used for the Uehling potential in the software. The integration over $u$ is left to be carried out numerically.\newline
The second correction adopted is the contribution of an electronic background charge. In this case the potential is found by solving a differential equation

\begin{equation}\label{s:theory:pot:bkgpot-ode}
	\frac{\partial}{\partial r}\left(r^2	\frac{\partial V_Q}{\partial r}\right) = 4\pi r^2 \rho_{bkg}(r)
\end{equation}

, which comes by simple application of Gauss's law in spherical coordinates.

\subsection{Transition probabilities}\label{S:theory:trans}

To compute the transition probabilities between two states in presence of an electromagnetic field we need to compute the dipole matrix element between them. For spontaneous emission between two states $a$ and $b$, in atomic units, we have \cite{payne1955}

\begin{equation}\label{s:theory:trans:transprob}
W_{ab} = \frac{4}{3}\frac{E_{ab}}{c}|\braket{a}{|\vec{\alpha}|j_0(kr)}{b}|^2
\end{equation}

with $E_{ab}$ transition energy, the wave vector has magnitude $k=E_{ab}/c$, and $\vec{\alpha}$ is the vector of Dirac matrices $\alpha_x, \alpha_y, \alpha_z$.
If we consider the wavefunction written as seen in eq. \ref{s:theory:dirac:fullpsi}, 
then it follows that each individual matrix element can be written as

\begin{align}\label{s:theory:trans:matel-alpha-j0}
\braket{a}{\alpha_ij_0(kr)}{b} = & i\braket{k_a\mu_a}{\sigma_i}{-k_b\mu_b}\int_{0}^{\infty}j_0(kr)P_aQ_b dr - \\
& i\braket{-k_a\mu_a}{\sigma_i}{k_b\mu_b}\int_{0}^{\infty}j_0(kr)P_bQ_a dr \nonumber
\end{align}

with the $\sigma_i$ being the Pauli matrices, keeping in mind that

\begin{equation}\label{s:theory:trans:alphamats}
\alpha_i = \begin{bmatrix}
0 & \sigma_i \\
-\sigma_i & 0
\end{bmatrix}
\end{equation}

. It should be recalled that in eq. \ref{s:theory:trans:matel-alpha-j0} the minus sign appears because the part in the bra is the adjoint, and not merely the complex conjugate, of the wavefunction. Now, keeping in mind eq. \ref{s:theory:dirac:spsphharm} and \ref{s:theory:dirac:sphharm}, and shortening the Clebsch-Gordan symbols seen in eq. \ref{s:theory:dirac:clebschgordan} as $c_{k\mu}^s = c(l,\frac{1}{2};\mu-s, s)$, we have

\begin{equation}\label{s:theory:trans:matel-kmu}
\ket{k\mu} = \begin{bmatrix}
c_{k\mu}^{1/2} Y_{l,\mu-1/2} \\
c_{k\mu}^{-1/2} Y_{l,\mu+1/2} \\
\end{bmatrix}
\end{equation}

and therefore,

\begin{equation}\label{s:theory:trans:matel-sz}
\braket{k'\mu'}{\sigma_z}{k\mu} = \left(c_{k'\mu'}^{1/2}c_{k\mu}^{1/2}-c_{k'\mu'}^{-1/2}c_{k\mu}^{-1/2}\right)\delta_{ll'}\delta_{\mu\mu'}
\end{equation}

\begin{equation}\label{s:theory:trans:matel-sp}
\braket{k'\mu'}{\sigma_+}{k\mu} = c_{k'\mu'}^{1/2} c_{k\mu}^{-1/2}\delta_{ll'}\delta_{\mu+1\mu'}
\end{equation}

\begin{equation}\label{s:theory:trans:matel-sm}
\braket{k'\mu'}{\sigma_-}{k\mu} = c_{k'\mu'}^{-1/2} c_{k\mu}^{1/2}\delta_{ll'}\delta_{\mu\mu'+1}
\end{equation}

. If we also set for brevity

\begin{equation}\label{s:theory:trans:pq-ints}
J_{ab} = \int_{0}^{\infty}j_0(kr)P_aQ_b dr
\end{equation}

, and we consider that if to a certain $k$ corresponds an $l$, then to $-k$ corresponds $l-\mathrm{sgn}(k)$, then eq. \ref{s:theory:trans:matel-alpha-j0} reduces to

\begin{align}\label{s:theory:trans:matel-alphax-j0}
\braket{a}{\alpha_xj_0(kr)}{b} = & i\left[\left(c_{k_a\mu_a}^{1/2} c_{-k_b\mu_b}^{-1/2}\delta_{\mu_a\mu_b+1}
+ c_{k_a\mu_a}^{-1/2} c_{-k_b\mu_b}^{1/2}\delta_{\mu_a+1\mu_b}\right)
\delta_{l_al_b-\mathrm{sgn}(k_b)} J_{ab}  - \right. \\
&\left. 
\left(c_{-k_a\mu_a}^{1/2} c_{k_b\mu_b}^{-1/2}\delta_{\mu_a\mu_b+1}
+ c_{-k_a\mu_a}^{-1/2} c_{k_b\mu_b}^{1/2}\delta_{\mu_a+1\mu_b}\right)
\delta_{l_a-\mathrm{sgn}(k_a)l_b} J_{ba} 
\right] \nonumber
\end{align}

\begin{align}\label{s:theory:trans:matel-alphay-j0}
\braket{a}{\alpha_yj_0(kr)}{b} = & -\left[\left(c_{k_a\mu_a}^{1/2} c_{-k_b\mu_b}^{-1/2}\delta_{\mu_a\mu_b+1}
- c_{k_a\mu_a}^{-1/2} c_{-k_b\mu_b}^{1/2}\delta_{\mu_a+1\mu_b}\right)
\delta_{l_al_b-\mathrm{sgn}(k_b)} J_{ab}  - \right. \\
&\left. 
\left(c_{-k_a\mu_a}^{1/2} c_{k_b\mu_b}^{-1/2}\delta_{\mu_a\mu_b+1}
- c_{-k_a\mu_a}^{-1/2} c_{k_b\mu_b}^{1/2}\delta_{\mu_a+1\mu_b}\right)
\delta_{l_a-\mathrm{sgn}(k_a)l_b} J_{ba} 
\right] \nonumber
\end{align}

\begin{align}\label{s:theory:trans:matel-alphaz-j0}
\braket{a}{\alpha_zj_0(kr)}{b} = & i\left[\left(c_{k_a\mu_a}^{1/2} c_{-k_b\mu_b}^{1/2}-
 c_{k_a\mu_a}^{-1/2} c_{-k_b\mu_b}^{-1/2}\right)
\delta_{l_al_b-\mathrm{sgn}(k_b)} J_{ab}  - \right. \\
&\left. 
\left(c_{-k_a\mu_a}^{1/2} c_{k_b\mu_b}^{1/2}-
c_{-k_a\mu_a}^{-1/2} c_{k_b\mu_b}^{-1/2}\right)\delta_{l_a-\mathrm{sgn}(k_a)l_b} J_{ba} 
\right]\delta_{\mu_a\mu_b} \nonumber
\end{align}

. From these it becomes then possible to compute the transition probability as seen in Eq. \ref{s:theory:trans:transprob}. The transition rules are easily seen: only transitions with $\Delta l = \pm 1$ and $\Delta \mu = 0, \pm1$ are allowed. Given any two states, the integrals $J_{ab}$ can be computed numerically.\newline
While individual transition rates for different values of the magnetic quantum number $\mu$ can be computed, these are not in practice resolved, since the Hamiltonian we use leaves the energies degenerate in $\mu$. Therefore, when considering a transition between two states with total angular momentum $j_b$ and $j_a$, we actually compute an average transition rate

\begin{equation}\label{s:theory:trans:avg-rate}
    \left< W_{ab} \right > = \frac{1}{2j_a+1}\sum_{\mu_a}\sum_{\mu_b} W_{ab}
\end{equation}

, summing over all possible starting states and averaging over all possible $2j_a+1$ degenerate end states.

\section{Algorithms and implementation}\label{S:algo}

\subsection{Grid and boundary values}\label{S:algo:grid}

All calculations in \texttt{mudirac} make use of a logarithmic radial grid, so that $r(x) = r_0e^x$. In this way the radial Dirac equation becomes

\begin{align}\label{s:algo:grid:dirac-log}
\frac{\partial Q}{dx} &= kQ + r(x)\left(mc-\frac{E-V(x)}{c}\right)P \\
\frac{\partial P}{dx} &= -kP + r(x)\left(mc+\frac{E-V(x)}{c}\right)Q
\end{align}

. Grid parameters are chosen case-by-case based on the parameters of the atom and the quantum numbers and energy of the state that's being integrated. In particular, by default we have

\begin{equation}\label{s:algo:grid:r0}
    r_0 = \max\left(\frac{1}{Z\mu}, R\right)
\end{equation}

, namely, $r_0$ is chosen as the maximum between the expected radius of an ideal Schr\"{o}dinger 1s orbital for a system with nuclear charge $Z$ and reduced mass $\mu$ and the finite radius of the nucleus.\newline
The boundary conditions are found by solving eq. \ref{s:theory:dirac:dirac} analytically for the $r\rightarrow 0$ and $r\rightarrow \infty$ limits. For a value of the energy $E$, defining $K = \sqrt{m^2c^2-E^2/c^2}$), we find at infinity \cite{silbar2010}:

\begin{align}\label{s:algo:grid:bound-inf}
P(r) &\approx e^{-Kr} \\
Q(r) &\approx -\frac{K}{mc+E/c}P(r)
\end{align}

. At zero the situation is more complex, and differs depending whether we're considering a finite or point-size nucleus. For a point-size nucleus we have \cite{grant2009}

\begin{align}\label{s:algo:grid:bound-0-point}
P(r) &\approx r^{\gamma} \\
Q(r) &\approx -\frac{Z}{c(\gamma-k)}P(r)
\end{align}

with $\gamma = \sqrt{k^2-Z^2\alpha^2}$, while for a finite size nucleus we have for $k < 0$

\begin{align}\label{s:algo:grid:bound-0-finite-kneg}
P(r) &\approx r^{-k} \\
Q(r) &\approx \frac{3Z}{2cR(-2k+1)}r^{-k+1}
\end{align}

and for $k > 0$

\begin{align}\label{s:algo:grid:bound-0-finite-kpos}
P(r) &\approx r^{k+2} \\
Q(r) &\approx \frac{2cR(2k+1)}{3Z}r^{k+1}
\end{align}
.

\subsection{Energy error estimation for eigenvalue search}\label{S:algo:error}

The radial Dirac equation is solved by a shooting method, integrating each step with a fourth order Runge-Kutta (RK4) algorithm \cite{press1992}. When using a shooting method to integrate a boundary value problem, one applies the following iterative process

\begin{enumerate}
    \item start with a trial value for the eigenvalue, $E$;
    \item integrate the equation forward from zero and backwards from infinity, using that trial value, meeting halfway at the turning point (namely, the point $r_t$ such that $V(r_t) = E$);
    \item compute an error based on how well the left and right solutions match at $r_t$;
    \item choose a new value of $E$ to minimize the error and repeat from step 2 until a certain tolerance is reached.
\end{enumerate}

The important thing to decide is how to evaluate the matching described at step 3, and then how to optimise the energy. One might be tempted to simply look at how the values of the forward- and backward-integrated functions match at the turning point, e.g. $\delta P = P_f(r_t)-P_b(r_t)$, but that would be a mistake. The functions at this point can not be normalised, and so they are always known up to a constant term. Since this constant will be likely different between left and right integration (as it depends only on the boundary conditions), any absolute comparison is meaningless. Instead, it makes sense to consider the error

\begin{equation}\label{s:algo:error:fwd-bck-err}
    \delta y = y_f-y_b = \left. \frac{Q}{P}\right|_f - \left. \frac{Q}{P}\right|_b
\end{equation}

which only depends on the ratio of $Q$ and $P$. When $E$ is an eigenvalue of the Dirac equation, namely, when a solution has been found, we expect $\delta y = 0$; more in general we expect we can find this energy by applying a steepest descent search and using the derivative of the error in the energy

\begin{equation}\label{s:algo:error:derr-dE}
    \frac{\partial (\delta y)}{\partial E} = \frac{\partial y_f}{\partial E} - \frac{\partial y_b}{\partial E} = \zeta_f - \zeta_b
\end{equation}

to minimize it. Now, making use of eq. \ref{s:theory:dirac:dirac} we find that we can write a differential equation for $y$

\begin{equation}\label{s:algo:error:y-ode}
    y' = 2\frac{k}{r}y-g_+y^2+g_-
\end{equation}

where the apices indicate differentiation in $r$ and we defined for convenience

\begin{equation}\label{s:algo:error:g-pm}
g_\pm = \left(mc\pm\frac{E-V}{c}\right)
\end{equation}

. Differentiating in $E$ then we obtain

\begin{equation}\label{s:algo:error:zeta-ode}
    \zeta' =  2\left(\frac{k}{r}-g_+y\right)\zeta-\frac{1}{c}(y^2+1)
\end{equation}

(remember that $\partial g_+ / \partial E = - \partial g_- / \partial E = 1/c$). This differential equation can be easily integrated once we have a tentative $Q$ and $P$. The boundary conditions are found by differentiating the ones for $Q$ and $P$ and are

\begin{equation}\label{s:algo:error:zeta-bound-zero}
    \zeta(r) \approx 0 \qquad r \rightarrow 0
\end{equation}

\begin{equation}\label{s:algo:error:zeta-bound-inf}
    \zeta(r) \approx \frac{E}{c^2Kg_+} + \frac{K}{cg_+^2} \qquad r \rightarrow \infty
\end{equation}

. Integrating $\zeta$ forwards and backwards from the two extremes of the grid, we can then apply eq. \ref{s:algo:error:derr-dE} and use it to minimise the energy. The only problem is that $y$ has singularities in the poles of $P$, which causes practical difficulties in the integration of $\zeta$. This is easily fixed though if we consider an equation complementary to eq. \ref{s:algo:error:zeta-ode}, namely

\begin{align}\label{s:algo:error:eta-ode}
    \eta  &= \frac{\partial}{\partial E}\left(\frac{1}{y}\right) = -\frac{1}{y^2}\zeta \\
    \eta' &= -2\left(\frac{k}{r}+g_-\frac{1}{y}\right)\eta+\frac{1}{c}(\frac{1}{y^2}+1)
\end{align}

which can be integrated in all those parts of the domain in which $y$ grows too big. Performing a Runge-Kutta integration alternating between the two as needed it's possible to estimate $\zeta_f-\zeta_b$ and thus converge the energy to an eigenvalue. From eq. \ref{s:algo:error:fwd-bck-err} and \ref{s:algo:error:derr-dE} one can find that that

\begin{equation}
    \delta E \approx \delta y\frac{\partial E}{\partial (\delta y)} = \frac{y_f-y_b}{\zeta_f-\zeta_b}
\end{equation}

\begin{figure}[h]
    \centering
    \includegraphics[width=0.8\textwidth]{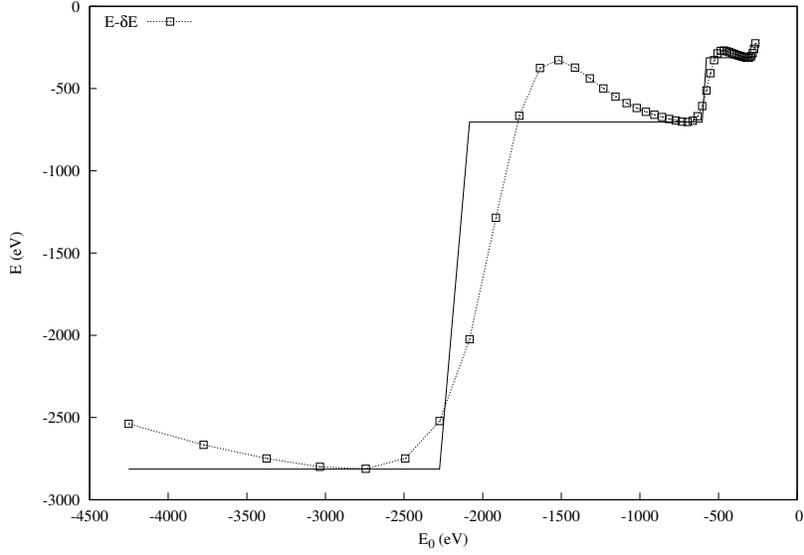}
    \caption{Predicted converged energy for various values of the starting guess in point-like muonic hydrogen. The solid line represents the final convergence energy corresponding to the state having the number of nodes found in the wavefunction at that value of the initial guess.}
    \label{s:algo:error:fig-scan-err}
\end{figure}

Figure \ref{s:algo:error:fig-scan-err} shows an example of the results produced by this algorithm. For a muonic hydrogen atom and a range of initial energy guesses $E_0$ it shows what the algorithm predicts as the converged energy, $E_0-\delta E$. If the linear approximation was correct, which of course tends to be true only when $E_0$ is close to an eigenvalue of the equation, then the figure would only show plateaus, with sudden jumps when the energy guess causes a new node to appear in the wavefunction, thus entering in the convergence basin of a different eigenvalue. These ideal plateaus, based on converged calculated energies and the number of nodes detected in the wavefunction, are shown as a solid line in figure \ref{s:algo:error:fig-scan-err}. The curve for $E_0-\delta E$ appears instead as an approximation to it, but one can see how in practice a series of updates $E_0' = E_0-\delta E$ does generally converge quickly to the appropriate eigenvalue. However, convergence tends to get more difficult when considering higher energy levels, where the gaps between successive shells shrink.

\subsection{Electronic background potential}\label{S:algo:bkg}

While in general the higher mass of the muon means its orbitals are much smaller than the electronic ones, and thus the contribution of the repulsion from electrons is negligible, there can be some overlap between some of the higher muonic orbitals and some of the lower electronic ones, which introduces a small correction to the energy. A full treatment of this correction would require solving the many-body problem for both the muon and all the electrons orbiting the atom, and at the moment has been considered unnecessarily complex for the scope of this software. Instead, the approach described by Tauscher et al. \cite{tauscher1978} has been used here.\newline
The scheme fills iteratively the electronic shells of the atom using the following rules:

\begin{enumerate}
    \item the shell with n = 1 is treated as seeing a nucleus of charge $Z-1$, to account for the shielding caused by the muon;
    \item each successive shell is treated as seeing a nucleus of charge $Z-1-q$, with $q$ the total number of electrons placed in lower shells.
\end{enumerate}

The orbitals used are the standard hydrogen-like solutions of the Dirac equation \cite{gordon1928}. The total background electronic density is then obtained as the sum of the individual density contributions from the radial parts of each of the filled orbitals. The exact density is calculated by using a user-defined electronic configuration.\newline
In order to compute the contribution to the potential of the electronic background, one needs to integrate eq. \ref{s:theory:pot:bkgpot-ode}. Let's assume the general case of a background charge $\rho(r)$ defined between the two limits $R_{in}$ and $R_{out}$. On a logarithmic grid the equation becomes

\begin{equation}\label{s:algo:bkg:bkgpot-ode-log}
    \frac{\partial^2 V}{\partial x^2} + \frac{\partial V}{\partial x}= 4\pi r^2 \rho(r)
\end{equation}

. This equation can not be integrated with the Numerov method that is most often used for second order differential equations due to the presence of a first order term. Instead, a third order scheme was used making use of the following finite differentiation approximations:

\begin{align}\label{s:algo:bkg:bkgpot-fdiff}
\left.\frac{\partial V}{\partial x}\right|_i &\approx \frac{1}{h}\left(\frac{11}{6}V_i -3V_{i-1} +\frac{3}{2}V_{i-2}-\frac{1}{3}V_{i-3} \right) \\
\left.\frac{\partial^2 V}{\partial x^2}\right|_i &\approx \frac{1}{h^2}\left(2V_i -5V_{i-1} +4V_{i-2}-V_{i-3} \right) \\
\end{align}

, setting the first three points of V with an approximation of constant charge for $r < R_{in}$ and shooting outwards for all the following ones. In general, the potential can be constructed piecewise at all values of $r$ as:

\begin{equation}\label{s:algo:bkg:bkgpot-piece}
V_{bkg}(r) = \begin{cases}
-\frac{Q}{r}\qquad & r > R_{out} \\
V_{grid}(r) -V_{grid}(R_{out}) - \frac{Q}{R_{out}} \qquad & R_{out} \geq r > R_{in} \\
\frac{2}{3}\pi r^2\rho(R_{in})-V_{grid}(R_{out}) - \frac{Q}{R_{out}} \qquad & r \leq R_{in}
\end{cases}
\end{equation} 

with $V_{grid}$ the potential integrated on the area of the grid in which $\rho$ is explicitly defined, and the convention $V_{bkg}(r) = 0$ when $r \rightarrow \infty$.

\section{Software details}\label{S:soft}

\texttt{mudirac} has been written to comply with the C++ 11 ISO standard \cite{cpp2011}. It makes use of the libraries Catch, for testing, and Aixlog, for logging, which are included with the distribution. It is compiled into an executable that can be ran as follows:

\begin{verbatim}
 $> mudirac input.in
\end{verbatim}

with the input file \texttt{input.in} (the name is arbitrary) containing all necessary parameters to determine the calculation. These include the atom type, the terms to include in the potential, and which spectral lines should be calculated. A list of possible keywords with their meaning and default values is presented in \ref{APP:keyw}.\newline
The software produces the following output files:

\begin{itemize}
    \item a \texttt{.log} and an \texttt{.err} file, containing regular logging of the program's operation and any errors/exceptions thrown during the run. The amount of information in the log can be changed by setting a verbosity parameter in the input file;
    \item a \texttt{.xr.out} file containing a full report of the energy (in eV) and transition probabilities (in $\mathrm{s}^{-1}$) for each requested spectral line, and excluding any lines for which calculations did not succeed or which were found to be forbidden transitions;
    \item optionally, a \texttt{.spec.dat} file containing a final simulated spectrum including all lines convoluted with Gaussian functions and with intensities proportional to their transition probability;
    \item optionally, a series of \texttt{.<name>.out} state files containing potential, $P$ and $Q$ components of the wavefunction for each computed state. The state names use the standard IUPAC X-ray spectroscopy convention \cite{jenkins1991};
    \item optionally, a series of \texttt{.<name1>.<name2>.tmat.out} files containing the details of the transition matrix between two given levels, namely, the transition rates between each of their respective degenerate states of different $m_j$.
\end{itemize}

Atomic units are used inside the code and for most input parameters (either energy or masses). The only exception are parameters for the writing of \texttt{.spec.dat} files, in which those that have units of an energy must be set in electronvolts.\newline
\texttt{mudirac} is open source software; it has been released publicly under the MIT License, and is available at the URL \url{https://github.com/muon-spectroscopy-computational-project/mudirac} as part of the tools produced by the Muon Spectroscopy Computational Project.

\section{Examples and tests}\label{S:ex}

\subsection{Hydrogen lines}\label{S:ex:H}

While \texttt{mudirac} was designed with muons in mind, it is possible to set the mass of the particle to arbitrary values, so solving for electronic atoms is not a problem. This is particularly important to test the quality of the transition probabilities, as it is not easy to find reference values for muonic atoms. As a first test, here we present the computed relativistic transition energies and probabilities for the hydrogen atom, compared to the well known and tabulated experimental values in the NIST Atomic Spectra Database.  Calculations with \texttt{mudirac} were run using a finite size nuclear model and the Uehling potential correction on. The lines are labelled  The results are shown in Table \ref{s:ex:H:compare} and show excellent agreement on both transition energies and probabilities.

\begin{table}[]
    \centering
    \begin{tabular}{|c|c|c|c|c|c|}
    \hline
        Line &  $E$ (eV) & $E_{NIST}$ (eV) & $A$ ($s^{-1}$) & $A_{NIST}$ ($s^{-1}$) & Ref. \\
    \hline
        K1-L2 & 10.19884 & 10.19886 & 6.2649e+08 & 6.2649e+08 & \cite{sansonetti2004,baker2008}\\
    \hline
        K1-L3 & 10.19889 & 10.19891 & 6.2648e+08 & 6.2648e+08 & \cite{sansonetti2004,baker2008}\\
    \hline
        L1-M2 & 1.88869 & 1.88869 & 2.2449e+07 & 2.2449e+07 & \cite{zhao1986, baker2008} \\
    \hline
        L1-M3 & 1.88870 & 1.88871 & 2.2448e+07 & 2.2448e+07 & \cite{zhao1986, baker2008} \\
    \hline
        L3-M4 & 1.88866 & 1.88867 & 1.0775e+07 & 1.0775e+07 & \cite{baker2008} \\
    \hline
        L3-M5 & 1.88866 & 1.88867 & 6.4651e+07 & 6.4651e+07 & \cite{hnsch1974, baker2008} \\
    \hline
        M1-N2 & 0.66104 & 0.66104 & 3.0652e+06 & 3.0652e+06 & \cite{baker2008} \\
    \hline
        M1-N3 & 0.66104 & 0.66105 & 3.0650e+06 & 3.0650e+06 & \cite{baker2008} \\
    \hline
    \end{tabular}
    \caption{Comparison between hydrogen line energies and transition probabilities computed with \texttt{mudirac} and reference values in the NIST database. Last column contains the papers from which the values were originally taken.}
    \label{s:ex:H:compare}
\end{table}

\subsection{Muonic atom lines}\label{S:ex:muA}

Here we present a few examples of comparisons of spectral lines for muonic atoms to values reported in the literature, in particular in \cite{borie1982}, \cite{kessler1975}, and the spectra from the database presented in \cite{zinatulina2018}. For the values found in \cite{borie1982}, a reference computed value is available, besides the experimental one. These computed values are extremely accurate, including also field theory effects beyond the Uehling potential, thus it's not surprising that the \texttt{mudirac} computed values differ slightly.
 Values from the other sources are all experimental. The comparisons can be seen in Table \ref{s:ex:muA:compare}.

\begin{table}[]
    \centering
    \begin{tabular}{|c|c|c|c|c|c|}
    \hline
         Line & \texttt{mudirac} & Ref. \cite{borie1982}, comp. & Ref. \cite{borie1982}, exp. & Ref. \cite{kessler1975} & Ref. \cite{zinatulina2018} \\
         \hline
         \textsubscript{12}Mg, L2-M4 & 56.390 & 56.392 & 56.392 & & \\
         \hline
         \textsubscript{12}Mg, L3-M5 & 56.215 & 56.216 & 56.216 & & \\
         \hline
         \textsubscript{26}Fe, K1-L3 & 1255.03 &  &  & & 1270.09 \\
         \hline
         \textsubscript{26}Fe, K1-M3 & 1520.25 &  &  & & 1531.69 \\
         \hline
         \textsubscript{26}Fe, K1-N3 & 1613.09 &  &  & & 1621.75 \\
         \hline
         \textsubscript{50}Sn, M4-N6 & 349.968 & 349.980 & 349.975 & & \\
         \hline
         \textsubscript{50}Sn, M5-N7 & 345.254 & 345.256 & 345.254 & & \\
         \hline
         \textsubscript{82}Pb, N6-O8 & 437.756 & 437.749 & 437.749 & & \\
         \hline
          \textsubscript{82}Pb, N7-O9 & 431.357 & 431.336 & 431.328 & & \\
         \hline
         \textsubscript{82}Pb, K1-L2 & 5728.6 &  &  & 5777.9 & 5778.9 \\
         \hline
         \textsubscript{82}Pb, K1-L3 & 5911.7 &  &  & 5962.7 & 5963.3 \\
         \hline
         \end{tabular}
    \caption{Computed and experimental energies for various muonic atom lines. Energies computed with \texttt{mudirac} make use of a finite size nucleus, Uehling correction to the potential, and an electronic background charge using the configuration of the atom with $Z-1$ (so, for example, the configuration for \textsubscript{12}Mg is that of neutral Na). Please note that the values from \cite{zinatulina2018} are manually extracted from images of spectra and thus may be less accurate than the others.}    \label{s:ex:muA:compare}
\end{table}

\subsection{Isotope effect}\label{S:ex:isotope}

\begin{figure}
    \centering
    \includegraphics[width=0.8\textwidth]{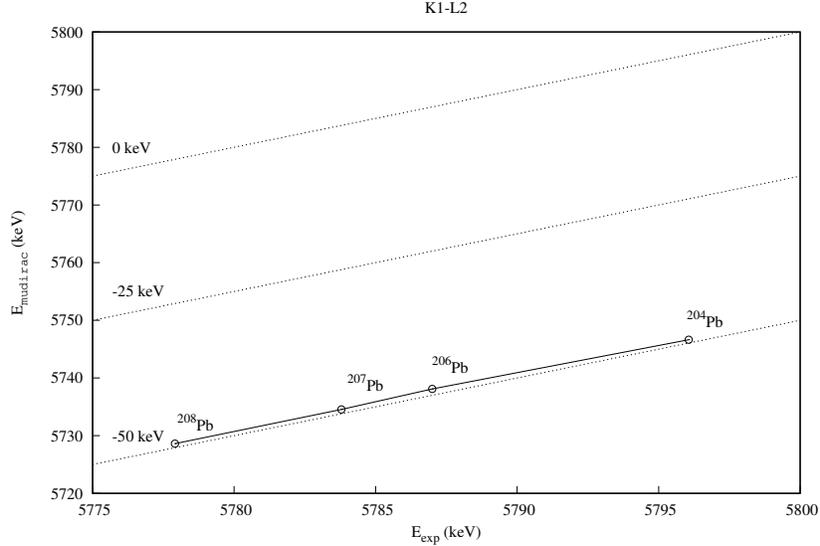}
    \caption{Experimental vs. computed energies for the K1-L2 transition in various isotopes of lead.}
    \label{s:ex:iso:K1L2}
\end{figure}

\begin{figure}
    \centering
    \includegraphics[width=0.8\textwidth]{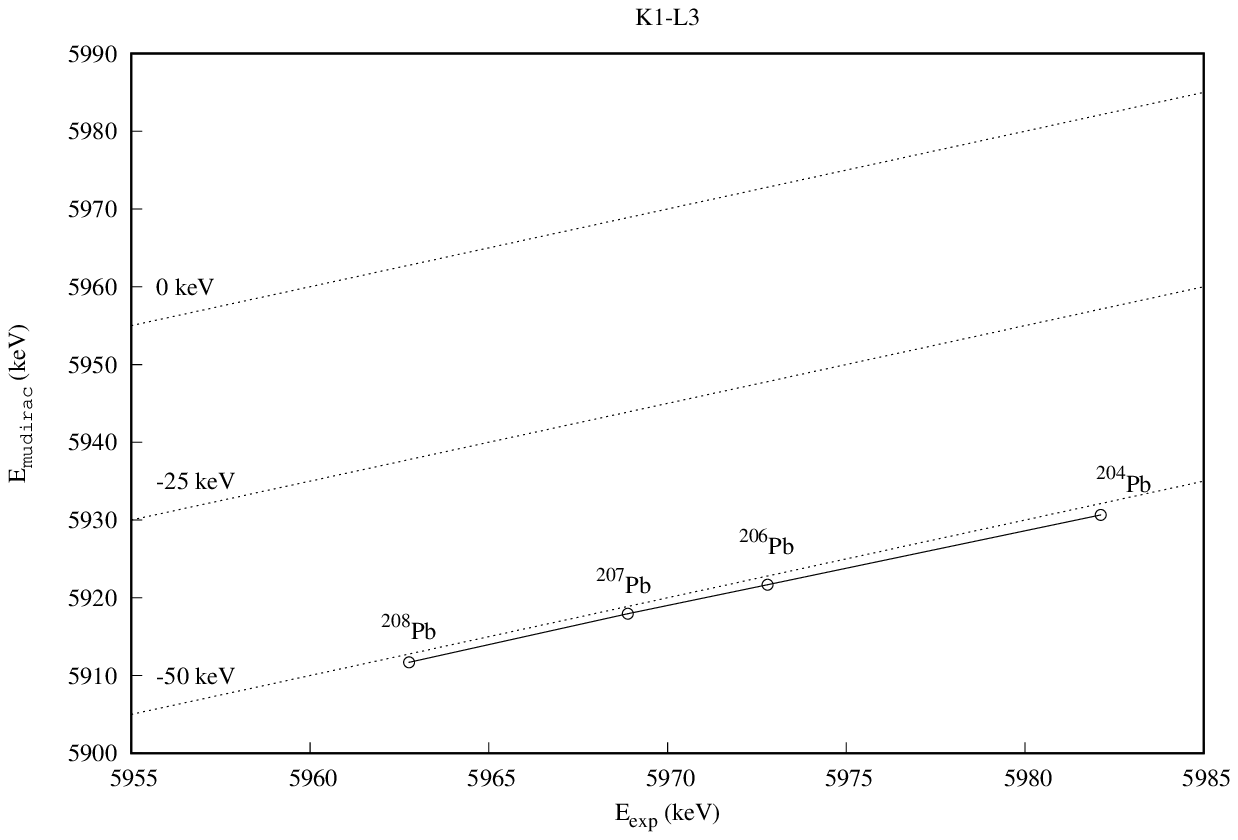}
    \caption{Experimental vs. computed energies for the K1-L3 transition in various isotopes of lead.}
    \label{s:ex:iso:K1L3}
\end{figure}

\begin{figure}
    \centering
    \includegraphics[width=0.8\textwidth]{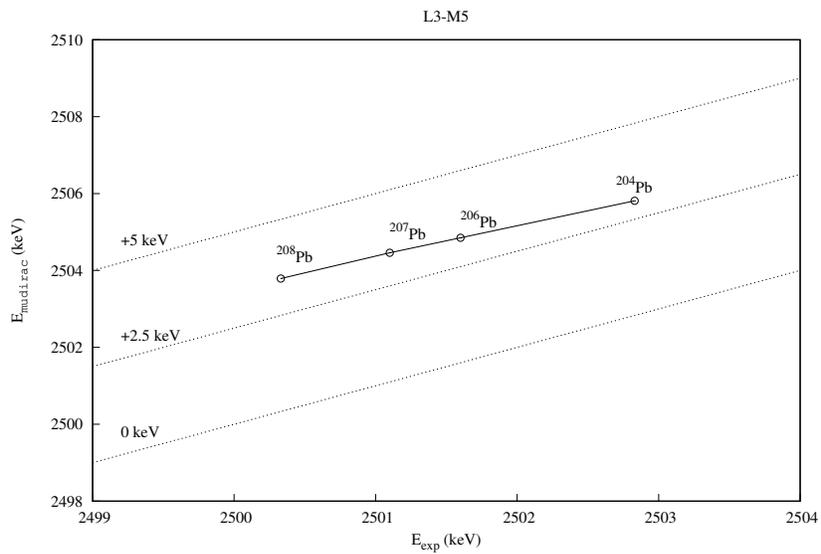}
    \caption{Experimental vs. computed energies for the L3-M5 transition in various isotopes of lead.}
    \label{s:ex:iso:L3M5}
\end{figure}

It has been shown \cite{ninomiya2019} that muonic X-ray analysis is also a good way to discern between isotopes of the same element, thanks to the effect of finite nuclear size and mass. Here we show the prediction made by \texttt{mudirac} for \textsuperscript{204}Pb, \textsuperscript{206}Pb, \textsuperscript{207}Pb and \textsuperscript{208}Pb, using as reference experimental data from the literature \cite{kessler1975}. Figures \ref{s:ex:iso:K1L2}, \ref{s:ex:iso:K1L3} and \ref{s:ex:iso:L3M5} show the predicted energies for the K1-L2, K1-L3 and L3-M5 transitions respectively for various isotopes of lead, compared to the known experimental values. The calculations were performed using a finite nuclear size, Uehling correction, but no electron screening. As it can be seen, the trend is reproduced very well, though there is a systematic error of around -50 keV for the K-L transitions, and of around +3 keV for the L-M one. This is likely due to one of the effects that have not been accounted for in the Hamiltonian \cite{borie1982}; it's interesting to note that it is in the same order of magnitude as the Uehling contribution to the line ($\sim$30 keV), which suggests it is not likely to be a higher order vacuum polarisation term. It is also clear that this effect dramatically affects the inner shells, as transitions between higher ones, even in Pb, are far more accurate (as seen in Table \ref{s:ex:muA:compare}. These will be considered for inclusion as the software develops, to guarantee higher accuracy. Nevertheless, the isotope effect is clearly well captured as the line shift from one isotope to another is much more accurate than the absolute value of the energies.

\subsection{Spectra comparisons}\label{S:ex:spectra}

\begin{figure}
    \centering
    \includegraphics[width=0.8\textwidth]{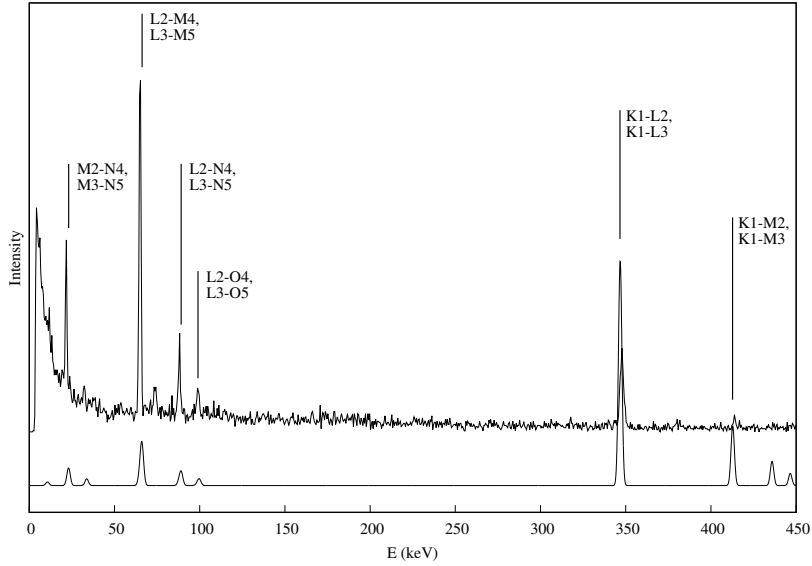}
    \caption{Experimental (top) and computed (bottom) X-ray spectrum for muonic aluminium.}
    \label{s:ex:spectra:Al}
\end{figure}

Finally, we consider a few example experimental spectra and compare them with \texttt{mudirac} computed results. The experimental spectra were acquired at ISIS with the RIKEN Port 4 instrument, and come in multiple resolutions, ranging from 2 to 0.5 keV between each point. The samples are pure metal standards. All calculations include a Uehling correction and the standard electronic configuration for the given element. Simulated spectra have been generated simply by using a Gaussian broadening of 1 keV. Energies are in keV, and intensities are arbitrary units. \newline
Figure \ref{s:ex:spectra:Al} shows a comparison for the spectrum of aluminium, up to 450 keV. The experimental spectrum has resolution of 0.5 keV. The computed one includes all lines involving states up to the O shell (n=5). One thing that can be noticed is how the intensity ratios between far apart group of lines are remarkably different between the experimental and computed spectrum; for example the L-M lines appear much higher than the K-L ones in the experiment, whereas that relationship is inverted in the computed spectrum. The major factor contributing to this is the fact that the sensitivity of the instrument used to acquire the spectrum isn't constant, but goes down as the energy increases, and the spectrum itself hasn't been corrected for this. If one focuses on individual groups of lines, such as the L-M, L-N, and L-O ones, one sees that experimental and computed spectra exhibit similar patterns, as the lines are concentrated in a small enough interval of energies that the sensitivity of the instrument remains nearly constant. Another thing to consider however is that the intensities in the computed spectrum are based only on the probability of a transition to take place. In reality, the intensity will be controlled also by the population of the upper state from which the decay takes place. In other words, the peak intensities of the computed spectrum would be accurate only if all states in the muonic atom were equally likely to be occupied. This is in practice not the case. In fact, the probability distribution in each of the states in the lower shells heavily depends on the cascade process and the starting angular momentum of the muon upon capture \cite{hufner1966, akylas1978}. The cascade can be computed given the transition rates, but one still needs to make assumptions about the initial angular momentum distribution. This is part of future plans for this project, but is currently not implemented in the software.

\begin{sidewaysfigure}
    \centering
    \includegraphics[width=0.9\textwidth]{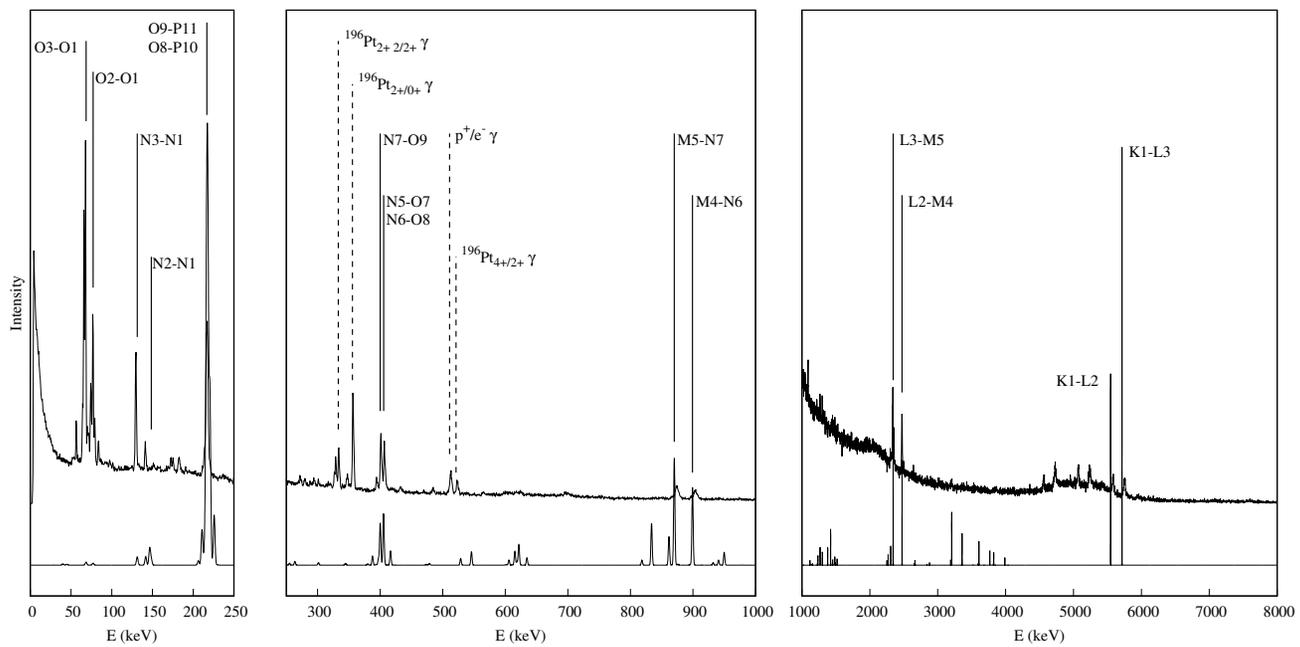}
    \caption{Experimental (top) and computed (bottom) X-ray spectrum for muonic gold.}
    \label{s:ex:spectra:Au}
\end{sidewaysfigure}

Figure \ref{s:ex:spectra:Au} shows a comparison between experiment and computation for gold, up to 8000 keV. The gold spectrum was computed up to the P shell (n=6). The comparison is split in different panels to highlight the details of the spectra at different scales; because the detector loses efficiency at high energy, the intensity has been rescaled between panels to better show the lines. The range up to 1000 keV uses data with a resolution of 0.5 keV; the 1000-8000 keV range uses data from a detector with a 2 keV resolution, which was the only one able to reach such high energies. 
The software allows us to identify various lines, from the K-L group at high energies (\textgreater 5000 keV) to the very low energy transitions in between states in the same shell (\textless150 keV). Interpretation of the spectrum, however, is in this case somewhat complicated by the fact that this vast range includes high energy phenomena that aren't only muon level transitions. Due to the high overlap between muon and nucleus, multiple nuclear reactions take place, resulting for example in the transmutation of gold to platinum by the reaction $\mu^-+p^+\rightarrow n + \nu $, with the excess energy ending up carried away by either the neutrino or the neutron, and the resulting nucleus being in an excited state that decays with emission of gamma rays. In particular, using the transition frequencies from the IAEA Nuclear Data database \cite{dunford1995}, the figure shows strong evidence that the reaction $^{197}\mathrm{Au}+\mu^-\rightarrow^{196}\mathrm{Pt}+n$ must be taking place, as several lines connected to the decay of excited states of $^{196}\mathrm{Pt}$ are identifiable (dashed lines in Figure \ref{s:ex:spectra:Au}). In addition to these, the 511 keV line corresponding to electron-positron annihilation can be seen. We can also see how the K1-L2 and K1-L3 lines, which have energies above 5 MeV, display single and double escape peaks at -511 and -1022 keV, a phenomenon related to the creation of electron-positron pairs and relative loss of energy in the measured photons. This is a well-known problem with germanium detectors in this energy range \cite{johnson1973}.

\begin{figure}
    \centering
    \includegraphics[width=0.8\textwidth]{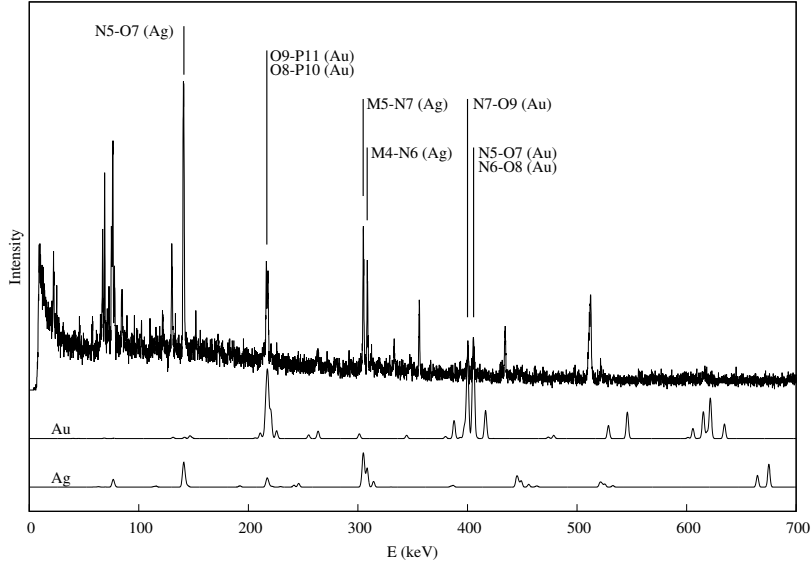}
    \caption{Experimental (top) X-ray spectrum for muons in a 70-30 Au-Ag alloy. Below are the computed spectra for muonic gold (middle) and silver (bottom).}
    \label{s:ex:spectra:AuAg}
\end{figure}

Finally, we consider the case of a two-element system; a 70-30 Au-Ag alloy. In this case our objective is to distinguish peaks belonging to each element, allowing for some basic elemental analysis. The Au computed spectrum was the same used in the previous section; the Ag one was computed with the same parameters, and again, up to the P shell (n=6). Figure \ref{s:ex:spectra:AuAg} shows the results. Here we focus on the low energies, as this is where the spectra are more rich in features, and thus interpretation is usually more difficult.

Some peaks are highlighted with their respective assignments. One interesting observation is that, again, intensities are dominated by capture and cascade mechanisms that obviously are not trivial; for example, despite the relative abundance of gold in the alloy, the M5-N7/M4-N6 doublet from silver is much higher than the N-O group from gold, running against the naive prediction based on transition probabilities. One must also pay attention to potential sources of ambiguity: for example, around 216 keV, one peak is assigned as the O9-P11/O8-P10 doublet from gold, but at the same frequency we can also see a peak in the silver spectrum. The structure of the peak, as well as its intensity, here lead us to believe it originates from gold; however it is hardly a foregone conclusion. Finally, another possible source of confusion must be considered in this low energy range: the lack of information about states above a certain energy. Computations were stopped at a given shell, which means that all transitions involving states above that are missing from our interpretation. As states get closer and closer in energy, going up in shells, more transition energies open up, leading to a nigh continuum of possible transitions in energy. This is further complicated by the fact that converging calculations for such higher states is also more difficult; in this example, the P10 and P11 states for Ag did not converge and the peaks they contribute to are missing from the spectrum. Going above this problem becomes only more severe, and affects our ability to properly interpret spectra in the low energy range. Improving these convergence properties is one of the future objectives for the development of the software.

\section{Conclusions}

A software to solve the radial Dirac equation specifically designed for muonic atoms has been produced and released for public use under an open source license. The software, called \texttt{mudirac}, is able to compute energies and transition rates for all the main X-ray spectral lines for muonic atoms across the periodic table, accounting for the most important effects such as the finite size of atomic nuclei and the Uehling correction. This provides a useful tool to help with the interpretation of muonic X-ray spectra for elemental analysis. The software has also been designed to easily accommodate future expansions such as the addition of more correction terms for its potential in order to achieve higher accuracy. In addition, future work will focus on using the software's predictions for transition rates in order to help interpret line intensities based on a model of the muonic cascade, as well as improving its ability to compute states in higher shells.

\section{Acknowledgements}

Thanks to Adrian Hillier, Dominik Jochym and Albert Bartok-Partáy for the useful discussions and contributions. The development of this software was funded by the Ada Lovelace Centre and was carried out in collaboration with the ISIS Muon Group.

\appendix

\section{Input keywords}\label{APP:keyw}

As explained in section \ref{S:soft}, \texttt{mudirac} takes a single input file, containing multiple lines with the format

\begin{verbatim}
    <keyword>: <value>
\end{verbatim}

. Here we list all the currently available keywords, divided by type, their purpose, and default values.

\subsection{String keywords}

These keywords take a string as value; invalid strings (e.g. a chemical symbol that doesn't correspond to a known element) will give rise to errors.

\begin{itemize}
    \item \texttt{element}: symbol of the element for the calculation. Determines the nuclear charge. Can be any symbol in the periodic table up to Z=111, Roentgenium (Rg). Default is H.
    \item \texttt{nuclear\_model}: model used to describe the nucleus. Can be POINT (point charge) or SPHERE (finite size, uniformly charged spherical nucleus). Default is POINT.
    \item \texttt{electronic\_config}: electronic configuration to use in order to describe the negative charge background. Can be a full string describing the configuration (e.g. \texttt{1s2 2s2 2p2}), an element symbol to represent the default configuration of that atom when neutral (e.g. \texttt{C}) or a mix of the two (e.g. \texttt{[He] 2s2 2p2}). Default is the empty string (no electrons).
    \item \texttt{ideal\_atom\_minshell}: for this shell, and all above it, treat the atom as a simple hydrogen-like point charge Dirac atom, using the known analytical solution and discarding all corrections. Mostly useful for debugging, or when very high shell states have difficulty to converge. The shell must use IUPAC notation (K $\implies$ n = 1, L $\implies$ n = 2, etc.). Default is the empty string (no ideal solutions used).
    \item \texttt{xr\_lines}: the transition or transitions for which energy and rates are desired. Each line must be expressed using the conventional IUPAC notation \cite{jenkins1991}. Multiple lines can be separated by commas. For example
    
    \begin{verbatim}
        xr_lines: K1-L2,K1-L3
    \end{verbatim}
    
    . In addition, colons can be used to indicate ranges of lines. The notation \texttt{K1:L3-M1} would compute the lines K1-M1, L1-M1, L2-M1 and L3-M1. Note that if some of these lines are forbidden by selection rules, they will simply be skipped. A double colon, like \texttt{K1:L3-K1:L3} would loop on both sides, and not count all repeated lines. 
\end{itemize}

\subsection{Boolean keywords}

These keywords can only have a value of TRUE or FALSE. In order to set them true, either the word `TRUE' or the letter `T' (regardless of case) work.

\begin{itemize}
    \item \texttt{uehling\_correction}: whether to turn on or not the Uehling correction. Default is FALSE.
\item \texttt{write\_spec}: if true, write a spectrum file using the transition lines found broadened with Gaussian functions. Other parameters can be specified with different keywords. Default is FALSE.
    \item \texttt{sort\_byE}: if true, print out the transitions sorted by energy instead than by shell. Default is FALSE.
\end{itemize}

\subsection{Floating point keywords}

These keywords accept a non-integer number. It can be written normally (e.g. 105.3) or in scientific notation (e.g. 1.053E2).

\begin{itemize}
\item \texttt{mass}: mass of the particle in atomic units (1 = mass of the electron). By default it's the mass of the muon, 206.7683.
\item \texttt{energy\_tol}: absolute tolerance for energy convergence when searching for eigenvalues. Iterations will stop once the energy change is smaller than this number, in atomic units. Default is 1E-7.
\item \texttt{energy\_damp}: a damping parameter used in steepest descent energy search to ease convergence. Used to multiply the suggested step $\delta E$ and make it smaller. Helps avoiding overshooting; fine-tuning it might help to converge difficult calculations, while making it bigger might make convergence faster in simple ones. Default is 0.5.
\item \texttt{max\_dE\_ratio}: maximum ratio between energy step, $\delta E$, and current energy $E$ in convergence search. If the suggested step exceeds this ratio times the guessed energy, it will be rescaled. This also serves as a measure to avoid overshooting and can be tweaked to get around cases of bad convergence. Default is 0.1.
\item \texttt{node\_tol}: tolerance parameter used to identify and count nodes in wavefunctions. Very unlikely to need any tweaking. Default is 1E-6.
\item \texttt{loggrid\_step}: step of the logarithmic grid. Default is 0.005.
\item \texttt{logggrid\_center}: center of the logarithmic grid in units of $a_0 = 1/(Zm)$. Default is 1.
\item \texttt{uehling\_lowcut}: low cutoff for Uehling potential, under which the radius will be considered 0. Default is 0.
\item \texttt{uehling\_highcut}: high cutoff for Uehling potential, over which the radius will be considered $r >> c$. Default is INFINITY.
\item \texttt{econf\_rhoeps}: charge density threshold under which the electronic charge background will be truncated and treated as zero. Default is 1E-4.
\item \texttt{econf\_rin\_max}: upper limit for the innermost radius of the electronic charge background grid. Default is -1 (no limit).
\item \texttt{econf\_rout\_min}: lower limit for the outermost radius of the electronic charge background grid. Default is -1 (no limit).
\item \texttt{spec\_step}: energy step for the simulated spectrum, in eV. Only has effect if \texttt{write\_spec = TRUE}. Default is 1E2 eV.
\item \texttt{spec\_linewidth}: Gaussian broadening width for the simulated spectrum, in eV. Only has effect if \texttt{write\_spec = TRUE}. Default is 1E3 eV.
\item \texttt{spec\_expdec}: exponential decay parameter $E_{dec}$ for a sensitivity function for the simulated spectrum, in eV. Multiplies the entire spectrum by a function $\exp(-E/E_{dec})$. Only has effect if \texttt{write\_spec = TRUE}. Default is -1 (no decay).
\end{itemize}

\subsection{Integer keywords}

Keywords that take an integer number as value.

\begin{itemize}
    \item \texttt{isotope}: which isotope of the element to consider. Important to determine the mass of the nucleus and its size. Default is -1, which means the most common isotope for the element will be used.
    \item \texttt{max\_E\_iter}: maximum number of iterations to perform when searching for the energy of a state. If exceeded, convergence will fail. Increase this value for slow convergences that are however progressing. Default is 100.
    \item \texttt{max\_nodes\_iter}: maximum number of iterations to perform when searching for a starting energy value that gives a state the expected number of nodes. If exceeded, convergence will fail. Should generally not need to be adjusted. Default is 100.
    \item \texttt{max\_state\_iter}: maximum number of iterations to perform when searching for a state. This loop encloses both node-based and energy-based search. Once a state is converged, the program checks again that it has the correct number of nodes. If it does not, the state is stored for future use and to provide an upper or lower limit to the energy of the searches and then the process is repeated. This number represents how much can the process be repeated before failing. Should not generally need to be adjusted. Default is 100.
    \item \texttt{uehling\_steps}: integration steps for the Uehling potential. Higher numbers will make the Uehling energy more precise but increase computation times. Default is 100.
    \item \texttt{xr\_print\_precision}: number of digits after the point to use when printing out energies and transition rates in the \texttt{.xr.out} file. Default is -1 (print as many as possible).
    \item \texttt{verbosity}: verbosity level. Going from 1 to 3 will increase the amount of information printed to the log file. Default is 1.
    \item \texttt{output}: output level. Going from 1 to 3 will increase the amount of files produced. Specifically:
    \begin{enumerate}
        \item will print out only the transition energies and rates in the \texttt{.xr.out} file;
        \item will print out also each of the states in a separate ASCII file as well as the transition matrices for each line;
        \item is reserved for future uses and currently has the same effect as 2.
    \end{enumerate}
\end{itemize}

%% New version of the num-names style
\bibliographystyle{model1-num-names}
\bibliography{mudiracpaper.bib}






% \end{thebibliography}

\end{document}